\documentclass{article}

\usepackage{natbib}

\usepackage{epsfig}

\usepackage{amssymb,amsmath}
\usepackage[T1]{fontenc}

\newcommand{\bmath}[1]{\mbox{\boldmath $#1$}}

\begin{document}


\title{Detection of additive outliers in Poisson INteger-valued AutoRegressive  time series}
\author{Maria Eduarda Silva\footnote{Corresponding author: mesilva@fep.up.pt} \\
Faculdade de Economia, Universidade do Porto, \\ Rua Dr  Roberto Frias,
  s/n, 4200 464 Porto, Portugal \\ Isabel Pereira\footnote{isabel.pereira@ua.pt}\\ Departamento de
   Matem\'atica, Universidade de Aveiro, \\ Campus Santiago, 3810 193 Aveiro,
Portugal \\ Center for Research and Development in Mathematics and Applications}


\date{December 2011}
\maketitle

\begin{abstract}
Outlying observations are commonly encountered in the analysis of time
series. In this paper the problem of detecting additive outliers in
integer-valued time series is considered. We show how  Gibbs
sampling can be used to detect outlying observations  in INAR(1)
processes. The
methodology proposed is illustrated using examples as well as an
observed data set.

\end{abstract}




\section{Introduction}
\label{introd}

This work considers a Bayesian approach to the problem of modelling a Poisson integer valued
autoregressive time series contaminated with additive outliers.

 It is
well known  that unusual observations and intervention effects
 often occur in data sets and can have adverse effects on model
identification and parameter estimation. In the
framework of Gaussian linear time series the problem of detecting and
estimating outliers and other intervention effects has been
investigated by several authors including \citet{Fox1972},
\citet{Tsay1986}, \citet{Changetal1988},
\citet{Chenetal1993} and \citet{Justeletal2001}, among others. However,  the problem of modelling outliers and other intervention
effects in the context of time series of counts has, as yet, received
little attention in the literature albeit its relevance for inference
and diagnostics.  Moreover, in this context additional motivation stems from the fact
that the usual techniques for outlier removal are not adequate since often
lead to non integer values.  In the framework of count time series it
is worth mentioning the work of
\citet{FokianosFried2010}  who
investigate the problem of modelling intervention effects in INGARCH
models and \citet{Barczyetal2010,Barczyetal2011}  who
consider CLS estimation of the parameters of an INAR(1) model
contaminated, at known time periods, with
innovational and additive outliers, respectively.

The well-known assertion of George Box that while
all models are wrong some are useful, motivates that we approach the
issue of modelling outliers in integer-valued time series focusing  on the  integer valued autoregressive
model of order 1. In fact,  this model introduced independently by
\citet{AlOshAlzaid1987}  
and \citet{Mckenzie1985} to model time series of counts, has been
extensively studied in the literature and applied to many real-world
problems including statistical process control, \citep{Weiss2007}
because of its simplicity and easiness of interpretation. 

\begin{figure}[h]
\begin{center}
\includegraphics[scale=0.5]{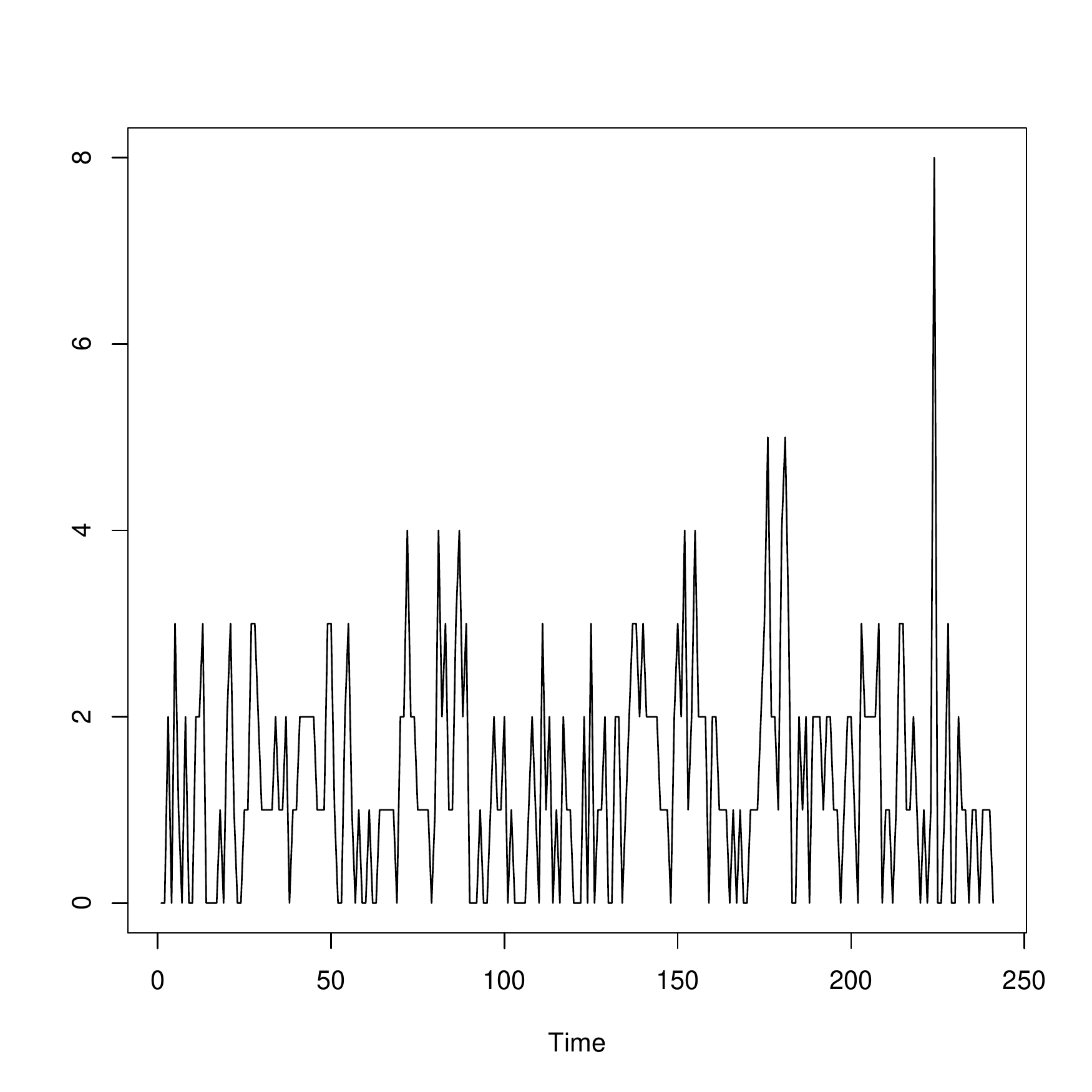}
\end{center}
\caption{Number of different IP adresses accessing the server of
the pages of the Department of Statistics of the University of
W\"urzburg between 10 am and 6 pm on 29 November  2005}
\label{ts_ip}
\end{figure}

 To motivate our approach, we represent in figure \ref{ts_ip} a
data set studied by \citet{Weiss2007}  concerning the number of
different IP addresses (approximately equivalent to the number of
different users) accessing the server of
the pages of the Department of Statistics of the University of
W\"urzburg in two-minute periods from  10 am to 6 pm on the 29th  November  2005, in a total of 241
observations. This time series is constructed from log data concerning
accesses to pages on the server.   \citet{Weiss2007} models the data with a Poisson INAR(1)
model and using statistical process control techniques finds an
outlying observation at time $t=224.$  As described by that author a
detailed 
analysis of the original log data showed  that all the eight accesses
at that time came from the  AOL browser that is known to supply
permanently new adresses within a small area. Therefore, it is not possible anymore
to infer the user from the IP address. It is
interesting to investigate if this observation can be explained by a
simple INAR(1) model and if the fit can be improved by the inclusion of
an additive outlier effect. 

Let $\{X_t\}$ be a Poisson INAR(1) process satisfying 
\begin{eqnarray}
 X_t & = & \alpha \circ X_{t-1} + e_t= \sum_{j=1}^{X_{t-1}}{\xi_{t,j}} + e_t,
\label{inar1}
\end{eqnarray}

with $(\xi_{k,j})$ a sequence of Bernoulli r.v.   with
mean  $\alpha \in [0,1]$ and  $\{e_t\},$ the arrival process,  a
sequence of i.i.d. Poisson variables $ e_t \sim
\mathcal{P}\!o(\lambda).$  When   additive outliers (AO) occur at
times $\tau_1, \ldots, \tau_k, $ with integer sizes
 $\omega_1, \ldots, \omega_k,$ $X_t$ is unobservable and it is assumed
 that the observed series $\{Y_t\}$ satisfies 

\begin{center}
$Y_t=X_{t} + \sum\nolimits_{i=1}^{k}{ I_{t,\tau_i} \omega_i},$
\end{center}

where $k\in \mathbb{N}$ is the number of outliers
and 
$I_{t,s}$ is an indicator function taking the value 1 if $t=s$ and 0
otherwise. Roughly speaking an additive outlier can
be interpreted as a measurement error or as an impulse due to some
unspecified exogenous source at time $\tau_i,$ $i=1,\ldots,k.$

Here we consider a Bayesian approach to the problem of Poisson
INAR(1) model specification in the presence of additive outliers.
Gibbs sampling provides estimates for the
probability of  outlier  occurrence at each time point leading to an 
effective outlier detection and accurate parameter
estimation. Bayesian approaches have been used to detect outliers in
ARMA models by \citet{Justeletal2001}  and in bilinear models by
\citet{Chen1997}.

The paper is organized as follows. Section \ref{INAR1ao} describes the
setup of additive outliers in INAR(1) models and explains the
procedure for outlier detection. Section \ref{Illustration}
illustrates the methodology on several sets of simulated data as well
as on a data set concerning the number of
different IP addresses accessing the server of
the pages of the Department of Statistics of the University of
W\"urzburg. Section \ref{Concl} concludes the paper.

\section{INAR$(1)$ models with additive outliers}
\label{INAR1ao}

Assume that the observed time series $Y_1,\ldots, Y_n$
is generated by 

\begin{equation}
Y_t = X_t+\eta_t\delta_t, \  \ 1 \leq
  t \le n 
\label{inar1contaminated}
\end{equation}
 where $X_t$ is a Poisson INAR(1) process satisfying (\ref{inar1}),
 $\delta_1, \ldots, \delta_n$ are 
 independent and identically distributed  Bernoulli variables with
 $P(\delta_t=1)=\epsilon$ and $\eta_1, \ldots, \eta_n$
 are independent random variables identically distributed as
 $Po(\beta).$  Also, $\delta_t$ and $\eta_t$ are independent for all
 $t$. This means that if $\delta_t=1$ the observation $Y_t$ is
 contaminated with  an AO of magnitude $\eta_t.$ Note that an outlier
 at time $t$ affects the model only at instants $t$ and $t+1.$

\subsection{Estimation procedure}

In this section we describe the Bayesian approach via Gibbs sampling
to estimate model (\ref{inar1contaminated}).  Assume that $Y_1=X_1,$ that is, there is no
 outlier in the first observation and let $\mathbf{Y}=(Y_1, \ldots,
 Y_n),$ 
$\bmath{\Theta}=(\alpha, \lambda),$ $\bmath{\delta}=(\delta_1,
\ldots, \delta_{n}),$ $\bmath{\eta}=(\eta_1,\ldots,\eta_n).$
Now we need to
derive the conditional posterior distributions of $\bmath{\Theta},$
$\bmath{\delta},$ $\bmath{\eta}$ and $\epsilon$. 

Conditioning on the first
observation the  likelihood of $\mathbf{Y}$ is given by

\begin{equation}
L(\bmath{\Theta},\bmath{\delta},\bmath{\eta},\epsilon)= \mathrm{e}^{-n\lambda} \ \prod_{t=2}^n \ \sum_{i=0}^{M_t} \
\frac{\lambda^{X_t-i}}{(X_t-i)!} \   \mathrm{C} ^{X_{t-1}}_{i}  \ \alpha^i \
(1-\alpha)^{X_{t-1}-i}
\label{likelihood}
\end{equation}

with $X_t=Y_t-\eta_t\delta_t$ and $M_t= \min{(X_{t-1},X_t)},$ $t=2,
\ldots,n.$

The prior distribution for the
 contamination parameter $\epsilon$ is   $\epsilon \sim$
 Be$(h,g),$ with expectation $E(\epsilon)=h/(h+g).$ Regarding the
 INAR(1) parameters $\alpha$ and $\lambda$ we choose for  prior
 distributions the conjugate of Binomial and Poisson, respectively and thus 
 $\alpha \sim$ Be$(a,b),$  $\lambda \sim$ Ga$(c,d) $
 \citep{SilvaSilvaPereiraSilva2005}. The set of hyperparameters 
$a,b,c,d,\beta,h,g$ are assumed known.

Let $\pi(\bmath{\Theta},\bmath{\delta},\bmath{\eta},\epsilon)$ denote
the prior distribution for
$(\bmath{\Theta},\bmath{\delta},\bmath{\eta},\epsilon).$ Then 

\begin{equation}
\pi(\bmath{\Theta},\bmath{\delta},\bmath{\eta},\epsilon) 
\propto \mathrm{e}^{-d \lambda}
\ \lambda^{c-1} \ \alpha^{a-1} \ (1-\alpha)^{b-1} \ \epsilon^{h-1} \
(1-\epsilon)^{g-1} \ \prod_{t=2}^{n} \mathrm{e}^{-\beta} \frac{\beta^{ \eta_t}}
  {\eta_t!}
\end{equation}

The posterior distribution of $\bmath{\Theta},$
$\bmath{\delta},$ $\bmath{\eta}$ and $\epsilon$ is then given by

\begin{eqnarray} 
\pi(\bmath{\Theta},\bmath{\delta},\bmath{\eta},\epsilon|\mathbf{y})   &
\propto       & L(\bmath{\Theta},\bmath{\delta},\bmath{\eta},\epsilon) \ 
\pi(\bmath{\Theta},\bmath{\delta},\bmath{\eta},\epsilon)   \nonumber \\
&  \propto &
\mathrm{e}^{-[d\lambda +n\beta]}
\ \lambda^{c-1} \ \alpha^{a-1} \ (1-\alpha)^{b-1} \ \epsilon^{h-1} \
(1-\epsilon)^{g-1}  \nonumber \\ 
 & & \ \ \frac{\beta^{\sum_{t=2}^{n} \eta_t}}{\prod_{t=2}^{n}
  \eta_t!} \ L(\bmath{\Theta},\bmath{\delta},\bmath{\eta},\epsilon)
\label{posteriorcompleta}
\end{eqnarray}

with $0<\alpha<1,$ $\lambda >0,$ $0 < \epsilon <1,$ and
$\eta_t=0,1,\ldots,$ $t=2,3, \ldots,n.$

The complexity  of the posterior marginals of $\bmath{\delta}$ and
$\bmath{\eta}$ suggest resorting to MCMC methods to implement the
Bayesian approach described above. 


The full  conditional posterior distributions for $\alpha$ and $\lambda$ 
are given by \citep{SilvaSilvaPereiraSilva2005}

\begin{equation}
\pi(\alpha|\mathbf{Y}, \lambda, \bmath{\delta},\bmath{\eta}, \epsilon
) \propto \alpha^{a-1} \ (1-\alpha)^{b-1} \ \prod_{t=2}^n
\ \sum_{i=0}^{M_t} T(t,i) \ \alpha^i \  (1-\alpha)^{X_{t-1}-i}
\label{palpha}
\end{equation}

with $T(t,i)=\frac{\lambda^{X_t-i}}{(X_t-i)!} \  \mathrm{C} ^{X_{t-1}}_{i}$

and

\begin{equation}
\pi(\lambda|\mathbf{Y}_n, \alpha, \bmath{\eta}, \epsilon,
\bmath{\delta}) \propto \lambda^{c-1} \mathrm{e}^{-(d+n)\lambda} \prod_{t=2}^n
\ \sum_{i=0}^{M_t} \ U(t,i) \ \lambda^{X_{t-i}}
\label{plambda}
\end{equation}

with $U(t,i)=\frac{1}{(X_t-i)!}  \mathrm{C} ^{X_{t-1}}_{i}  \ \alpha^i
(1-\alpha)^{X_{t-1}-i},$ respectively.

Now, with respect to the full conditional  distribution of
$\bmath{\delta}$ we reason as follows. For each $j=2, \ldots, n,$ $\delta_j| (\mathbf{Y},\alpha,\lambda,\bmath{\eta},
\epsilon,\bmath{\delta}_{(-j)}) \sim Ber(p_j),$ where
$\bmath{\delta}_{(-j)}$ denotes the vector $\bmath{\delta}$ with the
$j$th component deleted. Accordingly, we can write

\begin{equation}
p_j= P(\delta_j=1 |\mathbf{Y},\alpha,\lambda,\bmath{\eta},
\epsilon,\bmath{\delta}_{(-j)}) =  \frac{P(\delta_j=1
  ,\mathbf{Y}|\alpha,\lambda,\bmath{\eta}, \epsilon,\bmath{\delta}_{(-j)})
}{f(\mathbf{Y}|\alpha,\lambda,\bmath{\eta},
  \epsilon,\bmath{\delta}_{(-j)}) } 
\end{equation}

But 
\begin{align}
f(\mathbf{Y}|\alpha,\lambda,\bmath{\eta},
  \epsilon,\bmath{\delta}_{(-j)})= & P(\delta_j=1 |\ldots ) \
  f(\mathbf{Y}|\delta_j=1,\alpha,
  \lambda,\bmath{\eta},\epsilon,\bmath{\delta}_{(-j)}) \nonumber \\
& +P(\delta_j=0 |\ldots ) \ f(\mathbf{Y}|\delta_j=0,\alpha,
\lambda,\bmath{\eta},\epsilon,\bmath{\delta}_{(-j)}) \nonumber
\end{align}
 with  $P(\delta_j=1 |\ldots)=P(\delta_j=1 | \alpha,\lambda,
\eta_j,\epsilon)= \epsilon.$

Therefore

\begin{align}
p_j & 
 & =\frac{\epsilon f(\mathbf{Y}|\delta_j=1,\alpha,
  \lambda,\bmath{\eta}, \epsilon,\bmath{\delta}_{(-j)})}{\epsilon f(\mathbf{Y}|\delta_j=1,\alpha,
  \lambda,\bmath{\eta},
  \epsilon,\bmath{\delta}_{(-j)})+(1-\epsilon) f(\mathbf{Y}|\delta_j=0,\alpha,
  \lambda,\bmath{\eta}, \epsilon,\bmath{\delta}_{(-j)})}
\label{pdelta_1}
\end{align}


To compute $f(\mathbf{Y}|\delta_j=1,\alpha,
  \lambda,\bmath{\eta},
  \epsilon, \bmath{\delta}_{(-j)})$ first note that from (\ref{likelihood}) and the
Markovian property of the INAR(1) model the outlier at time $j$
affects the model for $t=j$ and $t=j+1.$ Then,

\begin{align}
f(\mathbf{Y}|\delta_j=1,\alpha,\lambda,\bmath{\eta},
\epsilon, \bmath{\delta}_{(-j)}) =  & f(X_j,
X_{j+1}|X_{j-1},\delta_j=1,\alpha,\lambda,\bmath{\eta}, \epsilon,\bmath{\delta}_{(-j)})
\nonumber \\ 
= & f(X_j|X_{j-1},\delta_j=1,\alpha,\lambda,\bmath{\eta},
\epsilon,\bmath{\delta}_{(-j)})  \nonumber \\
 \ \ \ \ \  &
\times f(X_{j+1}|X_{j},\delta_j=1,\alpha,\lambda,\bmath{\eta}, \epsilon,\bmath{\delta}_{(-j)})
\nonumber \\
\label{py}
\end{align}

with $f(X_t|X_{t-1})=  \mathrm{e}^{-\lambda} \ \sum_{i=0}^{M_t} \
\frac{\lambda^{X_{t}-i}}{(X_t-i)!} \ \mathrm{C}^{X_{t-1}}_i  \ \alpha^i \
(1-\alpha)^{X_{t-1}-i}$
and $M_t=\min(X_{t-1},X_t)$ as before. Moreover, if $\delta_j=1$
then $X_j=Y_j-\eta_j.$ Therefore

\begin{align}
f(X_j|X_{j-1},\delta_j=1,\alpha,\lambda,\bmath{\eta},
\epsilon,\bmath{\delta}_{(-j)}) 
 = & P(\alpha \circ X_{j-1}+
e_j=X_j|X_{j-1},\delta_j=1,...) \nonumber \\
  = & P(\alpha \circ X_{j-1}+
e_j=Y_j-\eta_j|X_{j-1},\delta_j=1,...) \nonumber \\
 = & \mathrm{e}^{-\lambda} \ \sum_{i=0}^{M_j}  \ \mathrm{C}_i^{X_{j-1}} \ \alpha^i
  \ (1-\alpha)^{X_{j-1}-i}
  \ \frac{\lambda^{Y_j-\eta_j-i}}{(Y_j-\eta_j-i)!} 
\label{fxjxjm1}
\end{align}

and

\begin{align}
 f(X_{j+1}|X_{j},\delta_j=1,\alpha,\lambda,\bmath{\eta}, \epsilon,\bmath{\delta}_{(-j)})
& = P(\alpha \circ X_{j}+
e_{j+1}=X_{j+1}|X_{j},\delta_j=1,\alpha,\lambda,\eta_j, \epsilon) 
\nonumber \\
& =  \mathrm{e}^{-\lambda} \  \sum_{i=0}^{M^*_j}  \ C_i^{Y_j-\eta_j} \ \alpha^i
  \ (1-\alpha)^{Y_j-\eta_j-i}
  \ \frac{\lambda^{X_{j+1}-i}}{(X_{j+1}-i)!} \label{fxjM1xj} 
\end{align}

with
$M^*_t=\min{(Y_t-\eta_t, \ X_{t+1})}.$

Similarly, if $\delta_j=0$ then $X_j=Y_j$ and therefore

\begin{align}
f(\mathbf{Y}|\delta_j=0,\alpha,\lambda,\bmath{\eta},
\epsilon,\bmath{\delta}_{(-j)})) =& 
\mathrm{e}^{-2\lambda} \nonumber \\ 
  & \prod_{t=j}^{j+1} \ \sum_{i=0}^{M_t} 
     C_i^{X_{t-1}}  \ \alpha^i \ (1-\alpha)^{X_{t-1}-i}\ 
\frac{\lambda^{X_t-i}}{(X_t-i)!}
\label{pdelta0}
\end{align}

To derive the conditional posterior distribution of $\bmath{\eta}$
note that if $\delta_j=0,$ no outlier at $t=j,$ there is no
information about $\eta_j$ except the prior. Then
$    \eta_j |(\mathbf{Y},\lambda,\alpha,\epsilon,\delta_j=0,\bmath{\eta}_{(-j)}) \sim Po(\beta).$ However,
if $\delta_j=1$ $\mathbf{Y}$ contains information about
$\eta_j. $ Therefore,

\begin{align}
\pi(\eta_j \
|\mathbf{Y},\lambda, \alpha,\epsilon,\delta_j=1,\bmath{\eta}_{(-j)}) &=
\nonumber \\
& \frac{\pi(\eta_j | \lambda,\alpha,\epsilon,\delta_j=1)  f(\mathbf{Y} |\lambda,\alpha,\epsilon,\delta_j=1,\eta_j )}{ \sum_{\eta_j=0}^\infty  \pi(\eta_j | \lambda,\alpha,\epsilon,\delta_j=1)  f(\mathbf{Y} | \lambda,\alpha,\epsilon,\delta_j=1,\eta_j )}    \nonumber \\ \nonumber \\
 &\propto  \ \mathrm{e}^{-\beta} \beta^{\eta_j}/ (\eta_j !)  \  f(X_j,X_{j+1} \ |
\eta_j, \delta_j=1, \alpha, \lambda,\epsilon), \nonumber \\
 & \eta_j=0,1,2,\ldots 
\label{peta}
\end{align}

with $f(X_j,X_{j+1} \ |
\eta_j, \delta_j=1, \alpha, \lambda,\epsilon)$
as given in (\ref{py}), (\ref{fxjxjm1}) and (\ref{fxjM1xj}) and $\bmath{\eta}_{(-j)}$ denoting the vector $\bmath{\eta}$ with the
$j$th component deleted.

Finally, the conditional posterior distribution for  $\epsilon$ depends
only on $\bmath{\delta}.$ Since the prior distribution of $\epsilon$
is $Be(h,g)$ the conditional posterior is given by

\begin{equation}
\epsilon|\mathbf{Y},\lambda,\bmath{\eta},\bmath{\delta} \equiv
\epsilon|\bmath{\delta} \sim
Be(h+k,g+n-1-k)
\label{pepsilon}
\end{equation}

where $k$ is the estimated number of outliers (number of $\delta_j$'s=1).

\subsection{Computational Issues}

We may use the full conditional distributions of 
$\alpha, \lambda$, $\bmath{\delta}=(\delta_2,
\ldots, \delta_{n}),$ $\bmath{\eta}=(\eta_2,\ldots,\eta_n)$ and
$\epsilon$  to draw a sample of a Markov chain which converges to the joint posterior distribution of the
parameters. In most  cases we can not generate directly from the full
conditionals. Since they are not log-concave densities we use Gibbs
methodology within Metropolis step. In particular the Adaptive
Rejection Metropolis sampling - ARMS  \citep{GilksBest1995} - is used inside
the Gibbs sampler. When the number of iterations is sufficiently
large, the Gibbs draw can be regarded as a sample from the joint
posterior distribution. Accordingly there are two key issues in the
successful  implementation of  this methodology: deciding the length of the
chain and the burn-in period and establishing the convergence of the
chain. We use a burn-in period of $M$ iterations and then
iterate the Gibbs sampler for a further $N$ iterations but retain only each $L$th
value. This thinning strategy reduces the autocorrelation within the chain.

Once  the posterior
probability of outlier
occurrence  at each time point, $p_j=P(\delta_j=1|\mathbf{Y}, \alpha,
\lambda,\bmath{\eta},\epsilon,\mathbf{\delta}_{(-j)})$ is estimated a
  cut-off point of 0.5 is used for detecting outliers, i.e. there is a
possible outlier when $\hat{p}_j > 0.5$. 

We now discuss the other relevant issue  in the proposed bayesian approach:  the choice of 
the hyperparameters for prior distributions.  Recall from the previous
section that $\alpha \sim$ Be$(a,b),$  $\lambda \sim$ Ga$(c,d).$ We
set   $a=b=c=d=0.001$ to use non informative prior distributions (Beta
and Gamma distributions with large variability). 
For the prior for $\epsilon \sim Be(h,g)$ we
  choose  $h=5,$ $g=95$ so that $E(\epsilon)=0.05$ to reflect the prior belief that  outliers
  occur occasionally with probability 0.05 for any time
  point. Regarding the parameter $\beta$ of the prior distribution for
  the size of the  outlier at time $t,$ $\eta_t \sim P(\beta)$ two
  approaches are pursued: an
  informative setup in which $\beta_{info}$ is set equal to  three  times the
  standard deviation of the 1-step-ahead prediction error and also a
  non-informative setup with 
$\beta_{ninfo}=30$ to reflect large variability.

\section{Illustration}
\label{Illustration}

In this section we illustrate the performance of the above procedure
with simulated data sets of 100 observations and the IP data example
of section \ref{introd}.

\subsection{Simulated data sets}

We consider time series simulated  from  several INAR(1) processes with $\alpha=0.15,0.5,
0.85$ and $\lambda=1,3,5$ with one and three
outliers of different sizes $\eta$ of order equal to three, five and seven times the
standard deviation. The times of the outliers are generated randomly.

 The Gibbs sampler used to obtain the Bayesian estimates is iterated $M+N=5000$ times and
the $L=5$th value of the last $N=2500$ iterations is kept, providing sample
sizes of 500 values from which the estimates are computed as the
sample means.

 The results are reported for $\beta_{ninfo}=30$
since they do not differ from those obtained with  $\beta_{info}.$

\begin{figure}[htb]
\begin{center}
\includegraphics[scale=0.55]{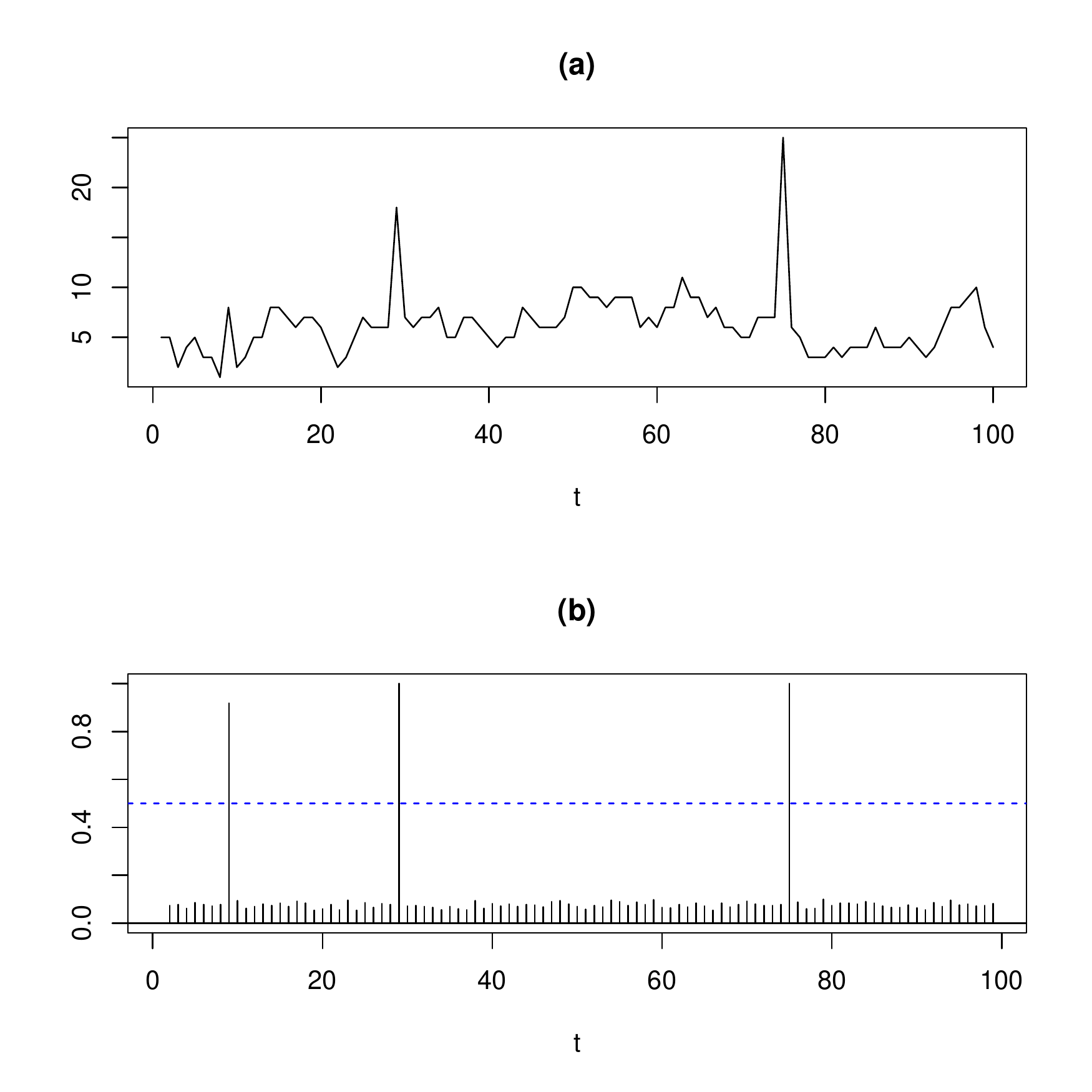}
\caption{(a) Simulated data with $\alpha=0.85,$ $\lambda=1,$ outlier at
  times $t=9,29,75$ with sizes $\eta=7,13,18,$ respectively; (b) posterior probability of
  outlier occurrence at
  each time point, $\hat{\eta}=7,12,19,$ respectively. }
\label{fig1}
\end{center}
\end{figure}

The results are illustrated  in  figure \ref{fig1} with simulated data from
the model with parameters $\alpha=0.85$, $\lambda=1$, outliers at $t=9,29,75$  with sizes
  $\eta=7,13,18,$ respectively.  Figure \ref{fig1} represents  the time series and
  the posterior probability of outlier occurrence for each time point,
  $\hat{p}_t.$  The Gibbs sampling successfully
  detects  the outliers
with estimated size of $\hat{\eta}_{9}=7,$ $\hat{\eta}_{29}=12$ and $\hat{\eta}_{75}=19.$ 

The results for all  the simulated models are summarized in table
\ref{table2} for series contaminated with 3 outliers.  The table
contains: the parameters $\alpha$ and $ \lambda$ used to generate the
series with outliers of size $\eta_S$ at times $S$, estimates for the parameters $\alpha$ and
$\lambda$ obtained by  conditional least
squares (assuming no outlier), \emph{Initial CLS} and  by Gibbs
sampling, \emph{Final Bayes}, the estimated
probability of outlier occurrence, \emph{Probability},  and the estimated outlier size for all the time
points for which that probability  is over the threshold 0.5, \emph{Final Bayes}. For
comparison purposes the table 
also presents the CLS estimates for the parameters $\alpha$ and
$\lambda$ after removing the effect of the detected outliers,
\emph{Final CLS}. The results presented in table \ref{table2} indicate that  the
procedure is able to detect additive outliers in INAR(1)
models. For  models with small variability ($\alpha$
and $\lambda$ small) small outliers are more difficult to detect. This
is illustrated for an INAR(1) model with
parameters $\alpha=0.15, \ \lambda=1$  and  outliers of size
$\eta_{50}=5$ and$\eta_{34}=7$ which are not detected. For
models with larger variability the outliers are correctly
detected even when their size is small (see figure 2). Moreover, the 
results in table  \ref{table2} illustrate the negative impact of the outliers on the
estimates of $\alpha$ and $\lambda$ (\emph{Initial CLS}).  It is
worthwhile noting that
the estimates obtained from Gibbs sampling (\emph{Final Gibbs}) and
the conditional estimates obtained removing the effect of the detected
outliers (\emph{Final CLS}) are, in general, similar. However, for
small $\alpha$ and for these particular simulated series, the Bayesian
estimates are biased which is a typical  behaviour for this range of
$\alpha$ values \citep{SilvaSilvaPereiraSilva2005}.

\begin{table}[htb]
\footnotesize{
\begin{tabular}{cccccc}
\hline
                 & & \multicolumn{3}{c}{Estimates} & \\ \cline{3-5}
Parameter & True & \emph{Initial} & \multicolumn{2}{c}{\emph{Final}} & \emph{Probability}
\\ \cline{4-5}
                 &        & \emph{CLS}   &  \emph{Bayes} & \emph{CLS} & \\ \hline \hline
$\alpha$ & 0.15        & 0.14       & 0.07& 0.17  &\\
$\lambda$ & 1        &    1.20    &1.27 &  1.05  &\\
$\eta_{34}$     &     7    &         & -- & & 0.15 \\ 
$\eta_{50}$  & 5 & & -- && 0.07 \\
$\eta_{63}$  & 9 & & 9 && 0.87 \\ \hline 
\hline
$\alpha$ & 0.15        & 0.09      &0.01 &0.03 &\\
$\lambda$ & 3        &    3.47    & 3.40& 3.40  &\\
$\eta_{34}$     &  9      &         & 11 & &  0.99\\ 
$\eta_{50}$  & 13 & & 13&& 0.99 \\
$\eta_{63}$  & 6 & & -- && 0.12  \\
 \hline 
\hline
$\alpha$ & 0.15        & 0.04      &0.32 &0.15 &\\
$\lambda$ & 5        &    6.35    & 4.0& 5.1 &\\
$\eta_{34}$     &  7      &         & -- & &  0.09\\ 
$\eta_{50}$  & 12 & & 13&& 0.96 \\
$\eta_{63}$  & 16 & & 18 && 0.99  \\
 \hline 
\hline
$\alpha$ & 0.5        &   0.22      & 0.41 & 0.37 &\\
$\lambda$ & 1         &  1.04     & 0.94 & 1.05 &\\
$\eta_{9}$     &  10      &         & 11& & 0.90 \\ 
$\eta_{27}$  & 4 & & -- && 0.01 \\
$\eta_{97}$  & 7 & &8&&0.81  \\
 \hline 
\hline
$\alpha$ & 0.5      &   0.23      &    0.59  & 0.57& \\
$\lambda$ & 3       &   4.72    &     2.28 & 2.39 &\\
$\eta_{9}$     &  17     &         &19 & & 0.99 \\ 
$\eta_{27}$  & 12 & &16&& 0.99\\
$\eta_{97}$  & 7 & & 10 &&  0.99 \\
 \hline 
\hline
$\alpha$ & 0.5 &  0.26   &   0.51     & 0.57 &\\
$\lambda$ & 5 & 7.04     &  4.30     & 3.87 &\\
$\eta_{9}$     &  10      &         &14 & & 0.91 \\ 
$\eta_{27}$  & 21 & &22&& 0.99\\
$\eta_{97}$  & 15 & & 17 && 0.99 \\
 \hline 
\hline
$\alpha$ & 0.85 & 0.37&0.86 & 0.80& \\
$\lambda$ & 3 & 1.27 & 2.62& 3.90&\\
$\eta_{9}$     &  31      &         & 29 & & 0.92 \\ 
$\eta_{29}$  & 13 & &10&& 0.99 \\
$\eta_{75}$  & 22 & & 22&&  0.99\\
 \hline 
\hline
$\alpha$    & 0.85 & 0.46     &0.85 & 0.85&\\
$\lambda$ & 5       & 17.55 & 4.60 & 4.66 &\\
$\eta_{38}$     &  40     &          & 37    &               & 0.92 \\ 
$\eta_{41}$    & 28       &          & 27    &       & 0.99 \\
$\eta_{78}$    & 17     &            & 20    &       &  0.99\\
 \hline 
\hline
\end{tabular} }
\caption{Results from Gibbs sampling in simulated
INAR(1) time series with parameters $\alpha$ and $\lambda,$ three
outliers each of size $\eta_S$ at time $S.$}
\label{table2}
\end{table}

\subsection{IP data example}

\begin{figure}[htb]
\begin{center}
\includegraphics[scale=0.65]{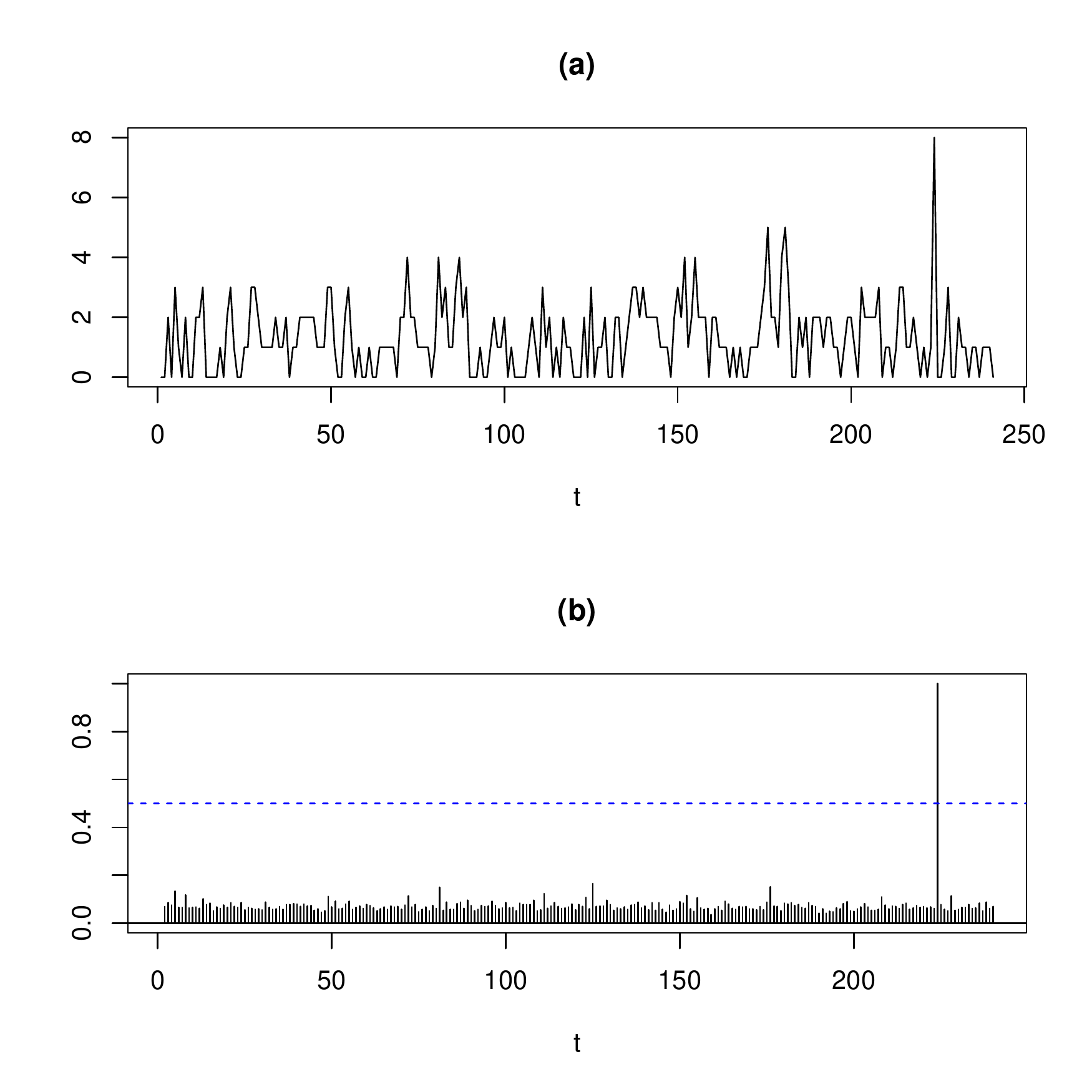}
\end{center}
\caption{Number of different IP adresses accessing the server of
the pages of the Department of Statistics of the University of
W\"urzburg {\bf(a)} and posterior probability of outlier occurrence at time $t$ for
  the IP data set {\bf(b)}.}
\label{delta_IP}
\end{figure}

\begin{figure}[htb]
\begin{center}
\includegraphics[scale=0.45]{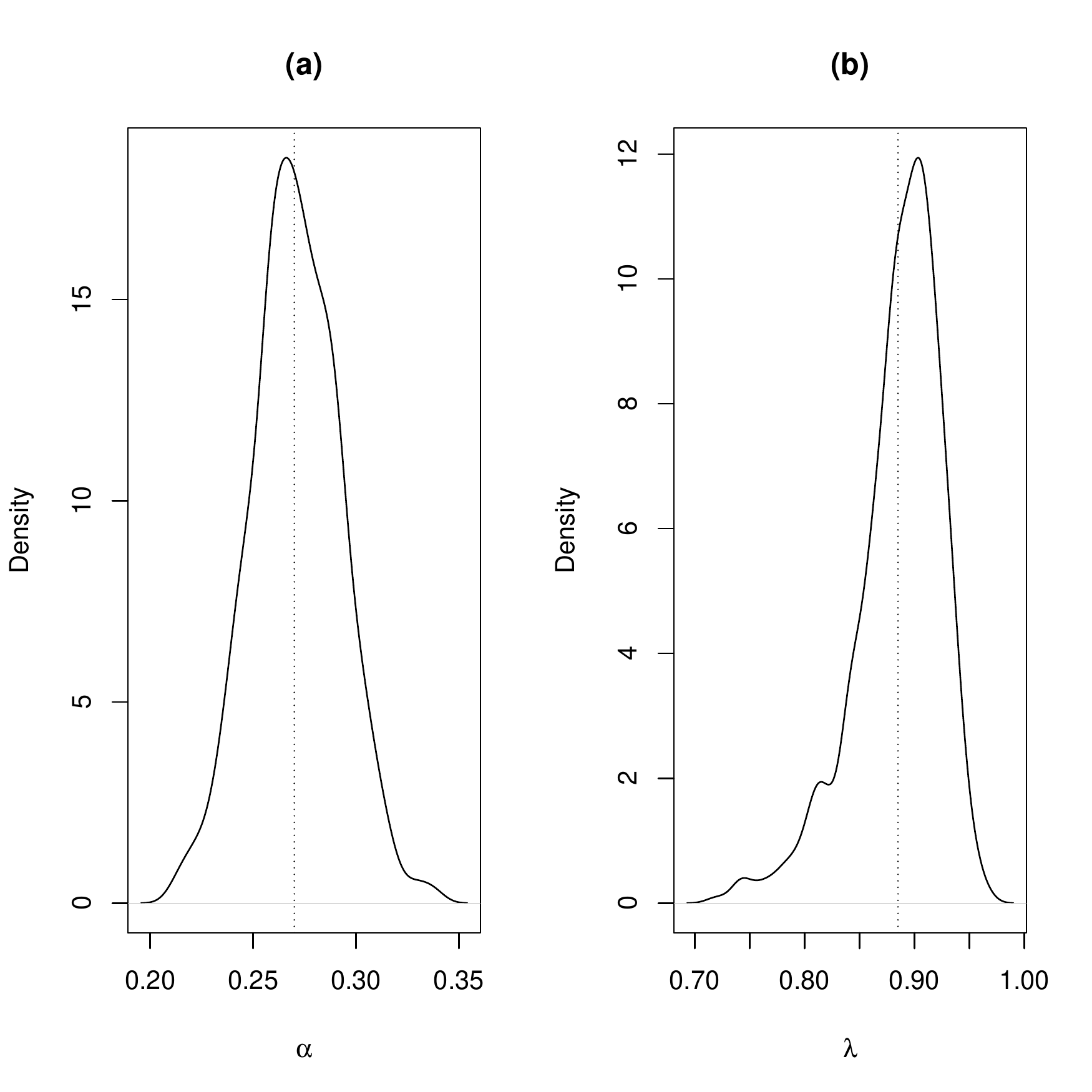}
\end{center}
\caption{Posterior distribution of $\alpha$ and $\lambda.$ The dotted
  lines represent the estimates  $\hat{\alpha}_{Bayes}=0.27$ and
$\hat{\lambda}_{Bayes}=0.89.$}
\label{posterior}
\end{figure}

Let us consider once again the motivating example of section
\ref{introd}, regarding the number of different IP addresses accessing  the server of the Department of Statistics
of the University of W\"urzburg on November 29th, 2005, between 10a.m. and 6p.m., represented in figure
\ref{ts_ip} \citep{Weiss2007}. The sample mean and variance of the series are
$\bar{x}=1.32, \hat{\sigma}^2=1.39.$ The autocorrelation and partial
autocorrelation functions indicate that a model of order one is
appropriate.  CLS estimates for
$\alpha$ and $\lambda$ are $\hat{\alpha}=0.22$ and
$\hat{\lambda}=1.03,$ respectively. 
The result of applying the proposed methodology is represented in  
figure \ref{delta_IP}(b) indicating the possible occurrence of
an outlier at time $t=224.$  The estimated size of the outlier is
 $\hat{\eta}=7.$ It is interesting to note
hat setting the time of the outlier to $t=224$ and using the results
from  \citet{Barczyetal2011} the CLS estimate for $\eta$ is
$\hat{\eta}_{CLS}=6.73.$ Removing the effect of the outlier at $t=224$ the mean and variance of
the resulting
series are  1.29 and 1.2, respectively. The autocorrelation and partial
autocorrelation functions still indicate that a model of order one is
appropriate. CLS estimates for
the parameters are now $\hat{\alpha}_{CLS}=0.29$ and
$\hat{\lambda}_{CLS}=0.91$ in accordance with the estimates obtained
from the Gibbs sampling, $\hat{\alpha}_{Bayes}=0.27$ and
$\hat{\lambda}_{Bayes}=0.89,$ whose posterior distribution is represented
in figure \ref{posterior}.


\section{Concluding remarks}

\label{Concl}
In this paper, the Gibbs sampling for detecting additive outliers
in Poisson INAR(1) time series is presented. We estimate the
probability that an observation is affected by an outlier. This procedure has
the advantage of identifying observations that may require further
scrutinizing. Note
that the hyperparameters of the prior distributions of the
outlier size and outlier occurrence probability, $\beta$ and
$\epsilon, $ respectively, are fixed but the same
methodology applies if they  are time dependent, $\beta_t$ and
$\epsilon_t,$ say.  Masking and swamping effects caused by patches of outliers may occur
depending on the size and relative position of the outliers within the
patch. The solution of this problem is being investigated. 

The extension of this methodology to models of
higher-order, INAR(p) $p>1,$ is possible. The mathematical
expressions are easily derived from the likelihood function. However,
the later  and consequently the
full conditional posterior distributions are highly
complex. Therefore, the implementation of the methodology for higher
order models requires additional computing effort. Since
applications of higher-order INAR models are scarce in the literature
this extension has not been considered in this work.

\section*{Acknowledgements}
\small{
  This work was supported by {\it FEDER} founds through {\it COMPETE}
  --Operational Programme Factors of Competitiveness (``Programa
  Operacional Factores de Competitividade'') and by Portuguese founds
  through the {\it Center for Research and Development in Mathematics
    and Applications} (University of Aveiro) and the Portuguese
  Foundation for Science and Technology (``FCT--Funda\c{c}\~{a}o para
  a Ci\^{e}ncia e a Tecnologia''), within project
  PEst-C/MAT/UI4106/2011 with COMPETE number
  FCOMP-01-0124-FEDER-022690.}





\section*{References}
\bibliographystyle{model2-names}
\bibliography{silva_pereira}

\begin{thebibliography}{14}
\expandafter\ifx\csname natexlab\endcsname\relax\def\natexlab#1{#1}\fi
\expandafter\ifx\csname url\endcsname\relax
  \def\url#1{\texttt{#1}}\fi
\expandafter\ifx\csname urlprefix\endcsname\relax\def\urlprefix{URL }\fi
\providecommand{\eprint}[2][]{\url{#2}}
\providecommand{\bibinfo}[2]{#2}
\ifx\xfnm\relax \def\xfnm[#1]{\unskip,\space#1}\fi
\bibitem[{Al-Osh and Alzaid(1987)}]{AlOshAlzaid1987}
\bibinfo{author}{Al-Osh, M.A.}, \bibinfo{author}{Alzaid, A.A.},
  \bibinfo{year}{1987}.
\newblock \bibinfo{title}{First-order integer-valued autoregressive (inar(1))
  process}.
\newblock \bibinfo{journal}{Journal of Time Series Analysis}
  \bibinfo{volume}{8}, \bibinfo{pages}{261--275}.
\bibitem[{Barczy et~al.(2010)Barczy, Ispány, Pap, Scotto and
  Silva}]{Barczyetal2010}
\bibinfo{author}{Barczy, M.}, \bibinfo{author}{Ispány, M.},
  \bibinfo{author}{Pap, G.}, \bibinfo{author}{Scotto, M.},
  \bibinfo{author}{Silva, M.E.}, \bibinfo{year}{2010}.
\newblock \bibinfo{title}{Innovational outliers in inar(1) models}.
\newblock \bibinfo{journal}{Communications in Statistics - Theory and Methods}
  \bibinfo{volume}{39}, \bibinfo{pages}{3343--3362}.
\newblock
  \eprint{http://www.tandfonline.com/doi/pdf/10.1080/03610920903259831}.
\bibitem[{Barczy et~al.(2011)Barczy, Ispány, Pap, Scotto and
  Silva}]{Barczyetal2011}
\bibinfo{author}{Barczy, M.}, \bibinfo{author}{Ispány, M.},
  \bibinfo{author}{Pap, G.}, \bibinfo{author}{Scotto, M.},
  \bibinfo{author}{Silva, M.E.}, \bibinfo{year}{2011}.
\newblock \bibinfo{title}{Additive outliers in inar(1) models}.
\newblock \bibinfo{journal}{Statistical Papers} ,
  \bibinfo{pages}{1--15}\bibinfo{note}{10.1007/s00362-011-0398-x}.
\bibitem[{Chang et~al.(1988)Chang, Tiao and Chen}]{Changetal1988}
\bibinfo{author}{Chang, I.}, \bibinfo{author}{Tiao, G.C.},
  \bibinfo{author}{Chen, C.}, \bibinfo{year}{1988}.
\newblock \bibinfo{title}{Estimation of time series parameters in the presence
  of outliers}.
\newblock \bibinfo{journal}{Technometrics} \bibinfo{volume}{30},
  \bibinfo{pages}{pp. 193--204}.
\bibitem[{Chen and Liu(1993)}]{Chenetal1993}
\bibinfo{author}{Chen, C.}, \bibinfo{author}{Liu, L.M.}, \bibinfo{year}{1993}.
\newblock \bibinfo{title}{Joint estimation of model parameters and outlier
  effects in time series}.
\newblock \bibinfo{journal}{Journal of the American Statistical Association}
  \bibinfo{volume}{88}, \bibinfo{pages}{pp. 284--297}.
\bibitem[{Chen(1997)}]{Chen1997}
\bibinfo{author}{Chen, C.W.}, \bibinfo{year}{1997}.
\newblock \bibinfo{title}{Detection of additive outliers in bilinear time
  series}.
\newblock \bibinfo{journal}{Computational Statistics and Data Analysis}
  \bibinfo{volume}{24}, \bibinfo{pages}{283 -- 294}.
\bibitem[{Fokianos and Fried(2010)}]{FokianosFried2010}
\bibinfo{author}{Fokianos, K.}, \bibinfo{author}{Fried, R.},
  \bibinfo{year}{2010}.
\newblock \bibinfo{title}{Interventions in ingarch processes}.
\newblock \bibinfo{journal}{Journal of Time Series Analysis}
  \bibinfo{volume}{31}, \bibinfo{pages}{210--225}.
\bibitem[{Fox(1972)}]{Fox1972}
\bibinfo{author}{Fox, A.J.}, \bibinfo{year}{1972}.
\newblock \bibinfo{title}{Outliers in time series}.
\newblock \bibinfo{journal}{Journal of the Royal Statistical Society. Series B
  (Methodological)} \bibinfo{volume}{34}, \bibinfo{pages}{pp. 350--363}.
\bibitem[{Gilks et~al.(1995)Gilks, Best and Tan}]{GilksBest1995}
\bibinfo{author}{Gilks, W.R.}, \bibinfo{author}{Best, N.G.},
  \bibinfo{author}{Tan, K.K.C.}, \bibinfo{year}{1995}.
\newblock \bibinfo{title}{Adaptive rejection metropolis sampling within gibbs
  sampling}.
\newblock \bibinfo{journal}{Journal of the Royal Statistical Society. Series C
  (Applied Statistics)} \bibinfo{volume}{44}, \bibinfo{pages}{pp. 455--472}.
\bibitem[{Justel et~al.(2001)Justel, Pe\~na and Tsay}]{Justeletal2001}
\bibinfo{author}{Justel, A.}, \bibinfo{author}{Pe\~na, D.},
  \bibinfo{author}{Tsay, R.}, \bibinfo{year}{2001}.
\newblock \bibinfo{title}{Detection of outlier patches in autoregressive time
  series}.
\newblock \bibinfo{journal}{Statistica Sinica} \bibinfo{volume}{11},
  \bibinfo{pages}{651--673}.
\bibitem[{McKenzie(1985)}]{Mckenzie1985}
\bibinfo{author}{McKenzie, E.}, \bibinfo{year}{1985}.
\newblock \bibinfo{title}{Some simple models for discrete variate time
  series1}.
\newblock \bibinfo{journal}{JAWRA Journal of the American Water Resources
  Association} \bibinfo{volume}{21}, \bibinfo{pages}{645--650}.
\bibitem[{Silva et~al.(2005)Silva, Silva, Pereira and
  Silva}]{SilvaSilvaPereiraSilva2005}
\bibinfo{author}{Silva, I.}, \bibinfo{author}{Silva, M.},
  \bibinfo{author}{Pereira, I.}, \bibinfo{author}{Silva, N.},
  \bibinfo{year}{2005}.
\newblock \bibinfo{title}{Replicated inar(1) processes}.
\newblock \bibinfo{journal}{Methodology and Computing in Applied Probability}
  \bibinfo{volume}{7}, \bibinfo{pages}{517--542}.
\newblock \bibinfo{note}{10.1007/s11009-005-5006-x}.
\bibitem[{Tsay(1986)}]{Tsay1986}
\bibinfo{author}{Tsay, R.S.}, \bibinfo{year}{1986}.
\newblock \bibinfo{title}{Time series model specification in the presence of
  outliers}.
\newblock \bibinfo{journal}{Journal of the American Statistical Association}
  \bibinfo{volume}{81}, \bibinfo{pages}{pp. 132--141}.
\bibitem[{Wei\ss(2007)}]{Weiss2007}
\bibinfo{author}{Wei\ss, C.H.}, \bibinfo{year}{2007}.
\newblock \bibinfo{title}{Controlling correlated processes of poisson counts}.
\newblock \bibinfo{journal}{Quality and Reliability Engineering International}
  \bibinfo{volume}{23}, \bibinfo{pages}{741--754}.

\end{thebibliography}

\end{document}